\documentclass[letterpaper, 10 pt, conference]{ieeeconf}
\IEEEoverridecommandlockouts
\usepackage{graphicx}
\usepackage{url}
\usepackage{svg}
\usepackage{amsmath}

\usepackage{amsthm}
\usepackage{amssymb}
\usepackage{dsfont}
\usepackage[caption=false, font=footnotesize]{subfig}
\usepackage{algorithmic}
\usepackage[ruled,vlined]{algorithm2e}

\usepackage{booktabs}
\usepackage{array}
\usepackage{float}
\usepackage{multirow}

\newtheorem{remark}{Remark}


\title{Pruning the Augmented Graphs of Convex Sets for\\ Scalable Joint Task and Motion Planning}
\author{Gael Luna and Tyler Summers
\thanks{The authors are with the Department of Mechanical Engineering at University of Texas at Dallas. Email: \texttt{tyler.summers@utdallas.edu}}
}

\begin{document}

\maketitle

\begin{abstract}

We present top-down and bottom-up approaches for solving large-scale instances of joint task and motion planning problems in Graphs of Convex Sets (GCS). Planning what tasks to perform and how to move between them can be encoded exactly as a Shortest Path Problem (SPP) using an Augmented GCS (AGCS), but this graph grows exponentially with the number of tasks, limiting prior work to only 11 tasks. Both of our approaches significantly improve scalability by carefully pruning the AGCS before it is constructed. The top-down approach replaces the complete graph with a sparse Delaunay graph that maintains a natural nearest-neighbor connectivity, reducing the number of vertices, edges, and subgraphs relative to the full AGCS. The bottom-up approach uses classical Traveling Salesman Problem (TSP) heuristics to create an initial ordering, then expands it within a fixed search window, pruning the AGCS to a bounded maximum width independent of the problem size. Both approaches can obtain optimal or near-optimal solutions in a fraction of the time required by the original AGCS. The bottom-up approach can solve instances with 1000 tasks in about two minutes. We additionally discuss lower bounds based on minimum 1-trees to quantify suboptimality of the proposed approaches.

\end{abstract}

\section{Introduction}

Path and motion planning are foundational problems in robotics \cite{lavalle2006planning}. Many such problems involve combinatorial optimization, where the optimal solution must be selected from a discrete set of candidates. Task planning also falls under the umbrella of combinatorial optimization. Path and motion planning address how the robot is going to get somewhere, while task planning answers where the robot needs to go for it to achieve a specific goal. Although intertwined, they are often solved separately or iteratively: first, robots are assigned to tasks, then the route is calculated for a given assignment. Joint task and motion planning produces superior performance but is computationally challenging.

A recent framework called Augmented Graphs of Convex Sets (AGCS) \cite{ you2026framework} solves task and motion planning jointly by encoding the task specifications directly into the graph structure. At its core, this framework is based on Graphs of Convex Sets (GCS) \cite{marcucci2024shortest,marcucci2025unified,marcucci2023motion}, which combines the discrete elements of graph theory with the continuous elements of convex optimization by assigning convex sets to vertices and convex functions as edge costs. GCS certifies optimal or near-optimal paths and trajectories for environments with obstacles. The AGCS converts a task and motion planning problem into a Shortest Path Problem (SPP) over an augmented graph consisting of copies of the environment's GCS.

\begin{figure}[!t]
  \centering
  \subfloat[]{\includegraphics[width=0.49\linewidth]{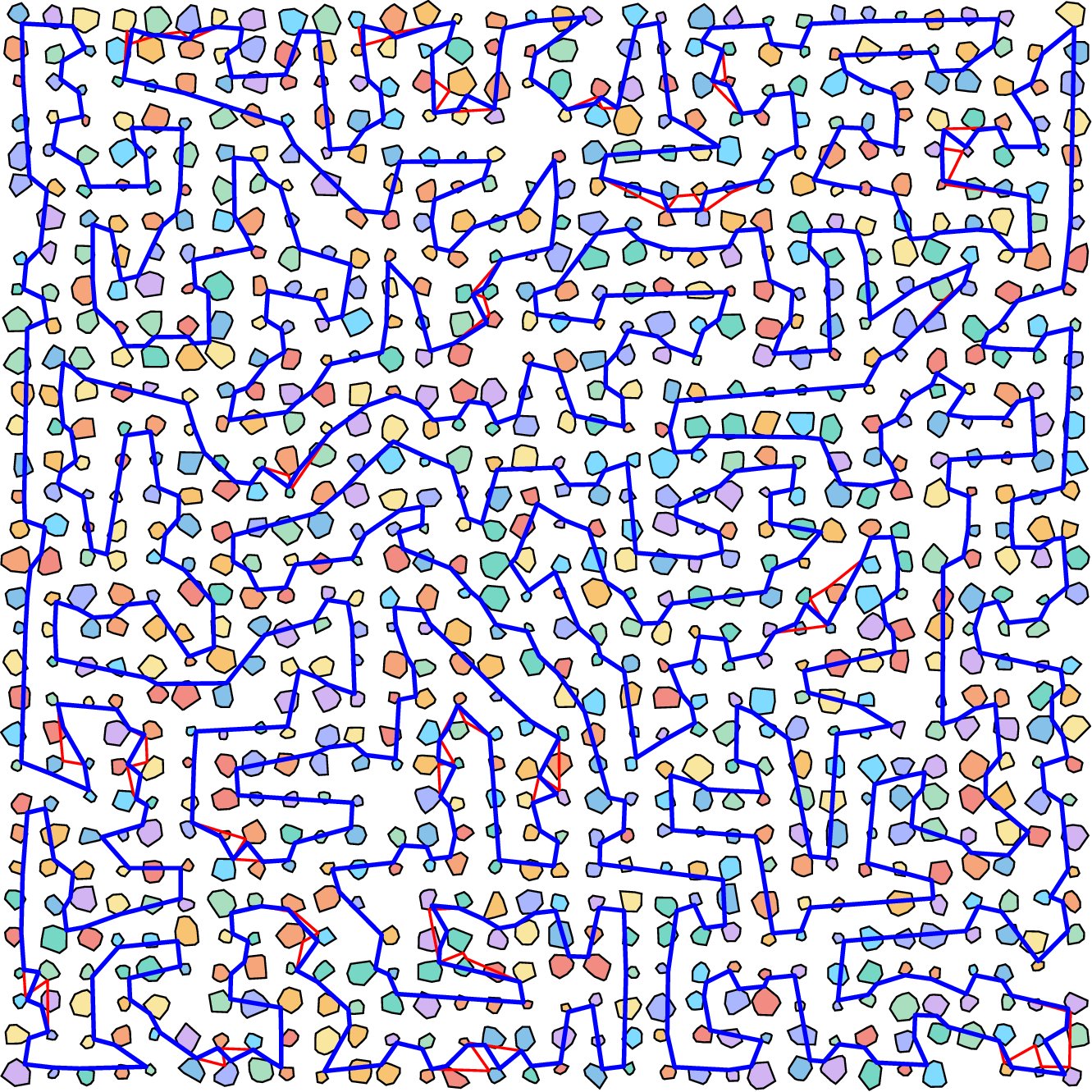}\label{fig:1000_sol}}
  \hfill
  \subfloat[]{\includegraphics[width=0.49\linewidth]{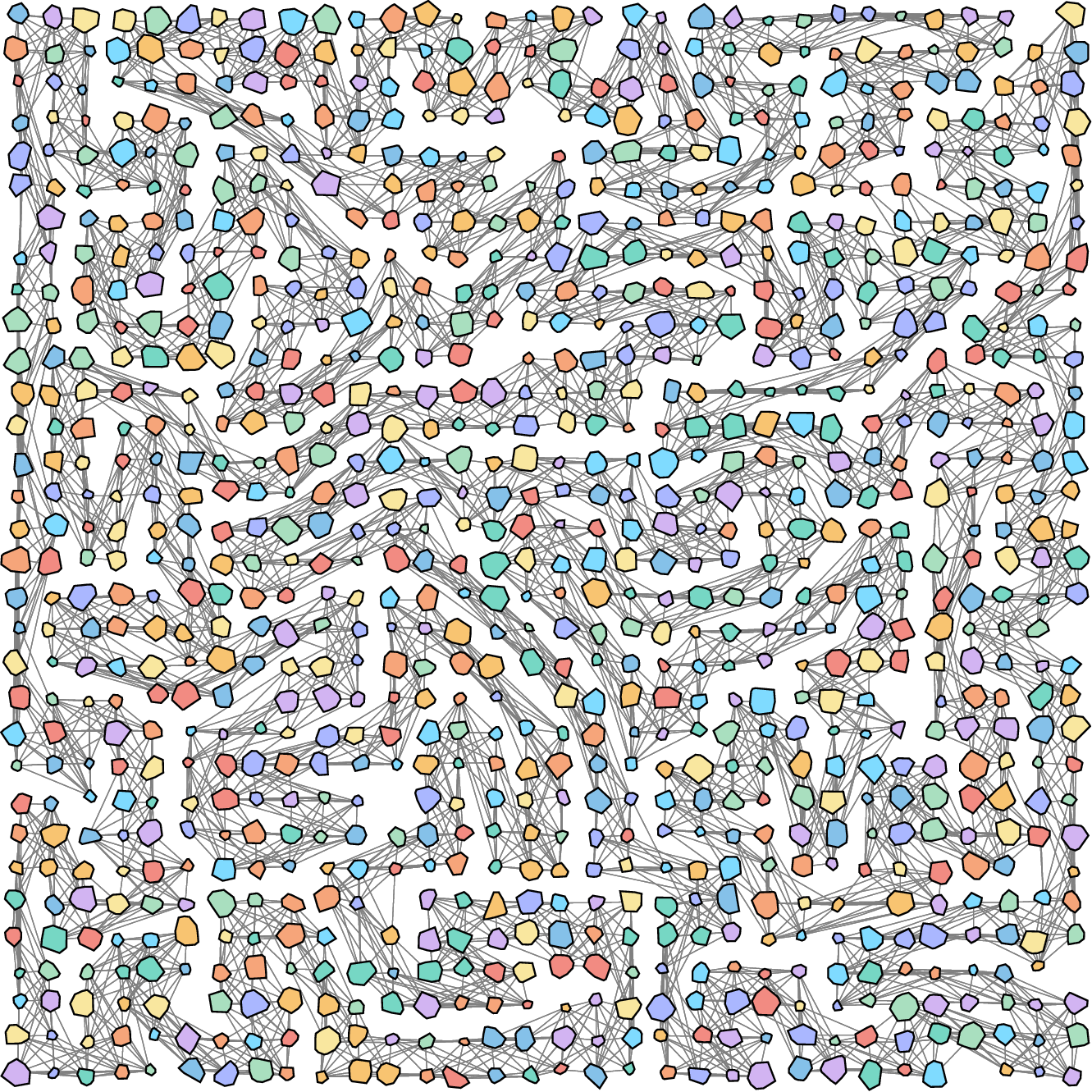}\label{fig:1000_window}}
  \caption{(a) An instance with 1000 tasks, placed in a grid like pattern, solved in 138.76 seconds using an AGCS with a maximum width of six. Red lines indicate the solution from the initial ordering, while blue lines indicate the final rounded solution using a pruned AGCS. (b) GCS based on an initial ordering and a search window of two.}
  \label{fig:large_tsp}
\end{figure}

Combined planning guarantees optimality, but at a computational cost that limited prior work \cite{you2026framework} to 11 tasks. This computational cost is determined by the maximum width of the AGCS, the largest number of subgraphs across all layers. In variation 2 of \cite{you2026framework}, where every task is reachable by every other task, the number of subgraphs is always $2^{n_{\mathcal{K}}}$, where $n_\mathcal{K}$ is the number of tasks. In this variation, the maximum possible width is always guaranteed. We refer to the regions being visited as \emph{task} regions, which are equivalent to the \emph{key regions} of the AGCS in \cite{ you2026framework}. By limiting the maximum width, we can scale to problems such as Figure \ref{fig:large_tsp} with 1000 tasks.

\begin{figure*}[htbp]
  \vspace*{4pt}
  \centering
  \includegraphics[width=0.85\textwidth]{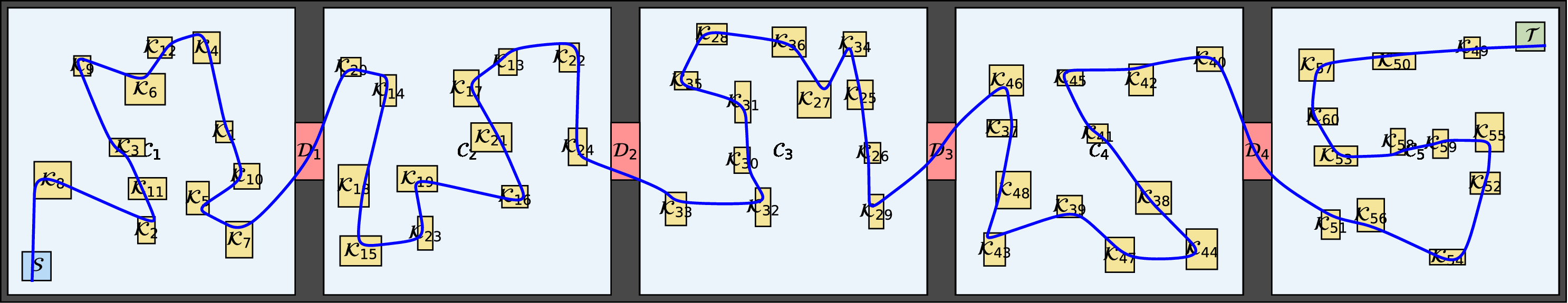}
  \caption{An instance with 60 tasks, blue is the source, green is the target, yellow are the tasks, red are the doors, blue-gray are the free-space partitions, and dark gray are the obstacles. This problem can be decomposed into five wayset-collection sub-problems: $(\mathcal{S}\to\{\mathcal{K}_{1},\dots,\mathcal{K}_{12}\} \to\mathcal{D}_{1})$, $(\mathcal{D}_{1}\to\{\mathcal{K}_{13},\dots,\mathcal{K}_{24}\}\to\mathcal{D}_{2})$, $(\mathcal{D}_{2}\to\{\mathcal{K}_{25},\dots,\mathcal{K}_{36}\}\to\mathcal{D}_{3})$, $(\mathcal{D}_{3}\to\{\mathcal{K}_{37},\dots,\mathcal{K}_{48}\}\to\mathcal{D}_{4})$, and $(\mathcal{D}_{4}\to\{\mathcal{K}_{49},\dots,\mathcal{K}_{60}\}\to \mathcal{T})$, where $\mathcal{S}$ and $\mathcal{T}$ are the source and target, respectively. Each door is considered a \emph{multi-task} door and treated as a source/target, for the respective sub-problem. These individual solutions can then be stitched together to form one solution. Solve time is 1.27 seconds.}
  \label{fig:traj}
\end{figure*}

The same bottleneck extends to special cases with precedence constraints, where task reachability is restricted by the environment. For example, Figure \ref{fig:traj} has 60 tasks, but due to the geometry not all 60 tasks are reachable from the source. In this example, the geometry forces five subproblems of 12 tasks each, which individually can be treated as a variation 2 problem. For this reason, the focus of this paper is on variation 2, specifically finding effective ways to trade optimality for computational efficiency.

\subsection*{Our contributions}
\textbf{Top-down pruning via Delaunay triangulation}. Replacing the complete graph with a sparse Delaunay graph keeps the natural nearest-neighbor connectivity while reducing the number of vertices, edges, and subgraphs relative to the full AGCS. On small instances where the full AGCS remains tractable, this approach recovers the optimal solution in a fraction of the time.

\pagebreak

\textbf{Bottom-up pruning via window search}. Classical TSP heuristics produce an initial ordering, which is then expanded within a fixed search window to produce an AGCS with a bounded maximum width, independent of the number of tasks. This approach can solve the same instances as the full AGCS magnitudes faster with minimal optimality gaps. Using this approach we can scale to problems with 1000 tasks such as Figure \ref{fig:large_tsp}. 

\textbf{Conservative lower bounds via minimum 1-trees}. We discuss why the convex relaxation of the AGCS becomes loose in special cases and why it becomes computationally expensive to compute as the problems scale. We then show how minimum 1-trees using bounded edge costs provide a conservative lower bound that remains scalable where the relaxation does not.

\subsection*{Related Work}
The main inspirations for the work in this paper are \cite{ you2026framework} for the AGCS, and \cite{helsgaun2000effective} for the use of minimum 1-trees in classical TSP heuristics. The TSP itself has some interesting and challenging variations \cite{ilavarasi2014variants} worth exploring in the GCS setting. One variation on the TSP which has been explored in depth in the GCS setting is the moving target variant \cite{philip2024moving,bhat2024moving,bhat2025moving}. In \cite{tang2026GHOST}, a traveling salesman-like problem in GCS was studied where the goal was to visit all regions in a partitioned environment. The Steiner TSP in GCS was studied in \cite{tang2026unified} where only select regions must be visited once, while the other regions are visited only if needed.

\section{Problem Formulation}
Consider a robot operating in an unbounded environment $\mathbf{R}^d$. The environment contains a set of tasks $\mathcal{K} = \{ \mathcal{K}_i \}_{i=1,\dots,n_\mathcal{K}}$. Each task $\mathcal{K}_{i}$ can be any convex set, but will be assumed to be a polytope with a given half-space representation:
\begin{equation} \nonumber
\begin{aligned}
\mathcal{K}_i &= \{ x \in \mathbf{R}^d \mid A_{\mathcal{K}_i} x \leq b_{\mathcal{K}_i} \}.
\end{aligned}
\end{equation}

Let $\mathcal{S}$ denote the source region and $\mathcal{T}$ the target region. Our goal is to find a trajectory $q$ over a time horizon $T$ that solves the following optimization problem:

\begin{equation} \label{optprob}
    \begin{aligned}
        &\underset{q}{\text{minimize}} \quad &&\alpha \int_0^T \| \dot q(t) \|_2 dt + \beta T \\
        &\text{subject to} && q(t) \models \phi(\mathcal{K}) \\
        & && \dot q(t) \in \mathcal{Q} \quad \forall t \in [0, T] \\
        & && q(0) = q_0 \in \mathcal{S}, \quad q(T) = q_T \in \mathcal{T}\\
        & && \dot q(0) = \dot q_0, \quad \dot q(T) = \dot q_T.
    \end{aligned}
\end{equation}

The objective is a weighted sum of the trajectory length and the time horizon with weights $\alpha$ and $\beta$, respectively. The first constraint is a temporal logic specification that requires each task to be eventually visited, but does not specify a visit order. The classical TSP in GCS additionally requires $\mathcal{S}=\mathcal{T}$; however, $\mathcal{S}$ and $\mathcal{T}$ are generally allowed to be distinct. The second constraint requires the velocity to be in the convex set $\mathcal{Q}$ for all time. The third constraint requires that the initial and terminal positions of the trajectory reside in the source and target regions. The fourth constraint controls the initial and terminal velocities.

The optimization problem \eqref{optprob} is infinite-dimensional, so candidate trajectories can be parameterized with piecewise B\'ezier curves defined by a finite set of control points. It is straightforward to include additional objective function terms and constraints that are convex in this parameterization. For example, penalties and constraints on higher-order derivatives promote additional smoothness of the trajectory $q$, like in Figure \ref{fig:traj}. Ensuring that the trajectory is differentiable a certain number of times facilitates the design of dynamically feasible trajectories for fully actuated and differentially flat systems.

\subsection{Linear Temporal Logic}
We use Linear Temporal Logic (LTL) \cite{pnueli1977temporal} to encode the TSP in GCS and wayset-collection specifications. LTL offers a general and powerful framework to specify complex tasks and constraints. Although LTL typically uses discrete time, here we use continuous time. An LTL formula can be composed from the following grammar
\begin{equation}
    \varphi ::= \top \mid p \mid \neg \varphi \mid \varphi_{1} \wedge \varphi_{2} \mid \varphi_{1} \vee \varphi_{2} \mid \varphi_{1} \mathcal{U} \varphi_{2} \mid \varphi_{1} \mathcal{R} \varphi_{2},
\end{equation}
by starting from a set of atomic propositions $AP$ with $p\in AP$ and recursively applying Boolean operators: $\neg$ (not), $\wedge$ (and), and $\vee$ (or), and the temporal operators $\mathcal{U}$ (until) and $\mathcal{R}$ (release). The until operator $\phi \mathcal{U} \psi$ is satisfied if $\phi$ remains true until $\psi$ becomes true. Additional temporal operators can be defined, such as eventually ($\mathcal{F} \varphi := \top \mathcal{U} \varphi $).

The TSP and wayset-collection specifications can be expressed using the eventually operator. In particular, we utilize a specific fragment of LTL of the form:
$$\varphi_{tsp} =  \bigwedge_{i=2}^{n_{\mathcal{K}}}
    \mathcal{F}(\mathcal{K}_i), \quad \varphi_{wayset} =  \bigwedge_{i=1}^{n_{\mathcal{K}}}
    \mathcal{F}(\mathcal{K}_i) \wedge \mathcal{F}(\mathcal{T})$$
which requires all tasks to be eventually visited but does not specify an ordering.

\subsection{Graphs of Convex Sets}\label{GCS}
The GCS framework \cite{marcucci2024shortest, marcucci2025unified} consists of a graph $\mathcal{G}(\mathcal{V}, \mathcal{E})$ with vertices $\mathcal{V}$ and edges $\mathcal{E}$. Each vertex $v \in \mathcal{V}$ is associated with a variable $x_v \in \mathbf{R}^{n_v}$, a convex set $\mathcal{X}_v \subseteq \mathbf{R}^{n_v}$, and a convex function $f_v : \mathbf{R}^{n_v} \rightarrow \mathbf{R} $. Each edge $e=(u,v) \in \mathcal{E}$ couples the vertex variables through a convex function $f_e : \mathbf{R}^{n_u + n_v} \rightarrow \mathbf{R}$ and a convex constraint set $\mathcal{X}_e \subseteq \mathbf{R}^{n_u + n_v}$. 

A general optimization problem on the GCS is

\begin{equation} \label{generalgcs}
    \begin{aligned}
        &\underset{H, \ x_v}{\text{minimize}} \quad &&\sum_{v \in \mathcal{V}_H} f_v(x_v) + \sum_{e=(u,v) \in \mathcal{E}_H} f_e(x_u, x_v) \\
        &\text{subject to} \quad && H = (\mathcal{V}_H, \mathcal{E}_H) \in \mathcal{H}, \\
        & && x_v \in \mathcal{X}_v,\quad \forall v \in \mathcal{V}_H, \\
        & && (x_u, x_v) \in \mathcal{X}_e, \quad \forall e = (u, v) \in \mathcal{E}_H, 
    \end{aligned}
\end{equation}

where the variables are the (discrete) subgraph $H$ with vertex set $\mathcal{V}_H \subseteq \mathcal{V}$ and edge set $\mathcal{E}_H \subseteq \mathcal{V}_H^2 \cap \mathcal{E}$ within an admissible subset of graphs $\mathcal{H}$, and the (continuous) variables $x_v$ for each vertex $v \in \mathcal{V}_H$.

\textbf{SPP in GCS}
For a given source vertex $\mathcal{S} \in \mathcal{V}$ and target vertex $\mathcal{T} \in \mathcal{V}$, a path $P$ is a sequence of distinct vertices that connects $\mathcal{S}$ to $\mathcal{T}$ via an edge subset $\mathcal{E}_P \subset \mathcal E$. The general problem \eqref{generalgcs} can be specialized to the SPP by taking $\mathcal{H}$ as the set of all paths, $\mathcal{P}$, from $\mathcal{S}$ to $\mathcal{T}$ in the graph $\mathcal{G}$. The SPP in GCS is stated as
\begin{equation} \label{spp_gcs}
    \begin{aligned}
        &\underset{P, \ x_v}{\text{minimize}} \quad &&\sum_{e=(u,v) \in \mathcal{E}_P} f_e(x_u, x_v) \\
        &\text{subject to} \quad && P =(\mathcal{V}_P, \mathcal{E}_P)\in \mathcal{P}, \\
        & && x_v \in \mathcal{X}_v,\quad \forall v \in \mathcal{V}_P, \\
        & && (x_u, x_v) \in \mathcal{X}_e, \quad \forall e \in \mathcal{E}_P
    \end{aligned}
\end{equation}
This problem seeks to simultaneously find a (discrete) path in the graph from the source vertex to the target vertex and the (continuous) values $x_v$ for each vertex that together minimize the total edge cost across all edges in the path. Although this problem is computationally difficult in general, a novel and tight mixed-integer formulation was developed in \cite{marcucci2024shortest}. This approach uses a network flow formulation of the SPP and exploits duality between perspective cones and valid inequality cones to convexify bilinear constraints that arise from the continuous vertex variables.

\section{Augmented GCS Construction}
The AGCS creates copies of the labeled GCS $\mathcal{G}(\mathcal{V}, \mathcal{E})$ from Section \ref{GCS} to encode which tasks have been visited. In \cite{you2026framework}, a correspondence was found between the Bellman-Held-Karp (BHK) \cite{held1962dynamic, bellman1962dynamic} algorithm and the AGCS, where the BHK algorithm uses binary subsets and the AGCS uses subgraphs, both representing the set of visited tasks. 

\subsection{Constructing the AGCS for the TSP in GCS}
\begin{figure}[htbp]
  \centering
  \includegraphics[width=0.95\linewidth]{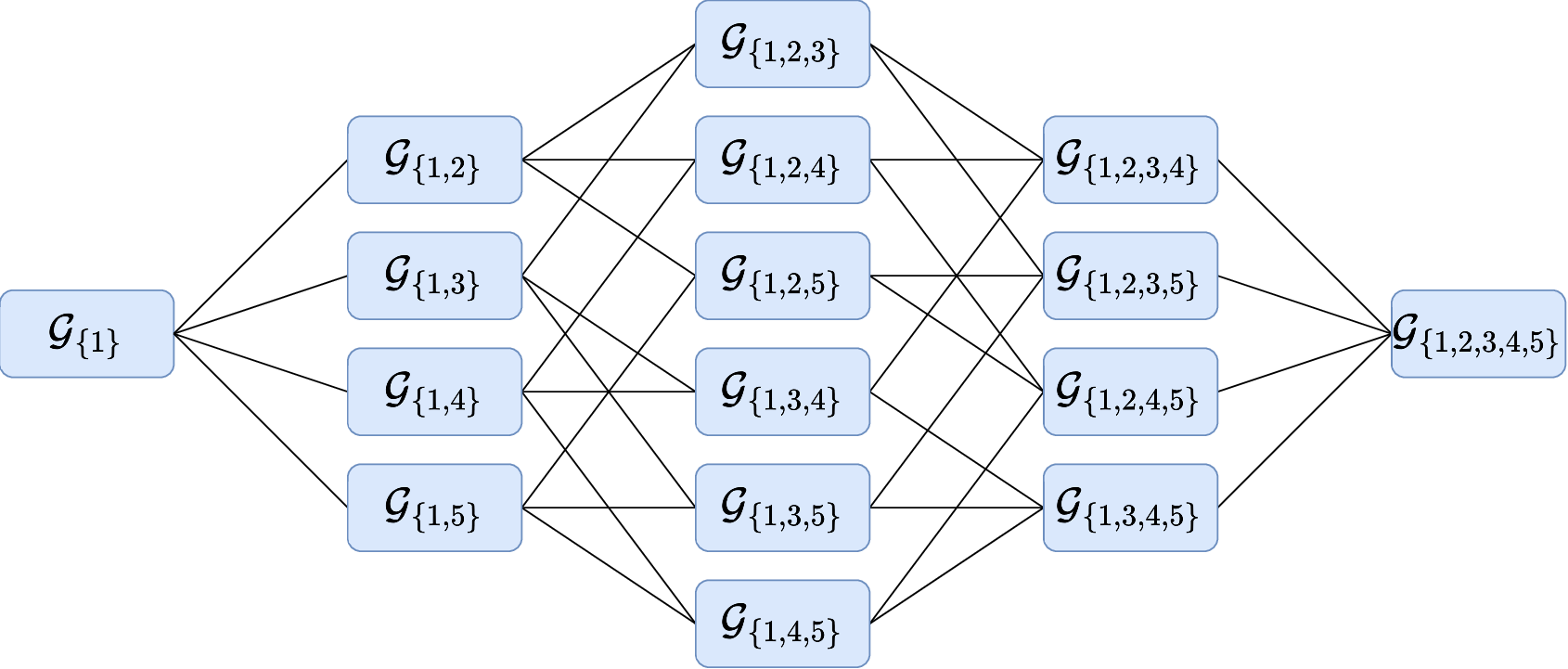}
  \caption{AGCS of a graph with five tasks and source $\mathcal{S}=\mathcal{K}_1$. Edges between subgraphs represent the visited task.}
  \label{fig:agcs_5}
\end{figure}

The AGCS is built recursively from the labeled graph $\mathcal{G}(\mathcal{V}, \mathcal{E})$ based on a source $\mathcal{S} \in \mathcal{V}$ whose corresponding convex set contains the initial condition. The AGCS consists of $n_\mathcal{K}$ layers, where each layer represents visited task subsets of a certain cardinality. Figure \ref{fig:agcs_5} shows an example where $n_{\mathcal{K}}=5$, which will be a useful reference to supplement the AGCS construction procedure below.

\textbf{Layer 1.} The first layer of the AGCS corresponds to the source. Since no visit order is specified, any task is a valid source. For simplicity, the source will be the task $\mathcal{K}_1$. This subgraph does not need to be modified and will contain the exact same vertex and edge sets as $\mathcal{G}(\mathcal{V}, \mathcal{E})$. Since the same graph will be used, each vertex is given a unique name based on the subgraph it belongs to, to differentiate vertices in one subgraph from those in another. For example, tasks $\mathcal{K}_1$ and $\mathcal{K}_2$ in subgraph $\mathcal{G}_{\{1\}}$ will be $\mathcal{K}_{1}^{\{1\}}$ and $\mathcal{K}_{2}^{\{1\}}$, respectively. 

\textbf{Layer 2.} The second layer of the AGCS corresponds to the subsets of two visited tasks, $B\in \mathcal{B}_2$ where $\mathcal{B}_2= \{ B \subseteq \mathcal{K} \mid |B| = 2, \mathcal{K}_1\in B\}$, every such subset must contain the source $\mathcal{K}_1$. For each subgraph in this layer, a copy of $\mathcal{G}(\mathcal{V}, \mathcal{E})$ is used, and like in the first layer, a suffix is added to each task. For example, $\mathcal{K}_1$ and $\mathcal{K}_2$ will be $\mathcal{K}_{1}^{\{1,2\}}$ and $\mathcal{K}_{2}^{\{1,2\}}$ in subgraph $\mathcal{G}_{\{1,2\}}$, respectively. A single directed edge is added from the parent subgraph in the first layer to the new subgraph in the second layer. For example, edge $\mathcal{K}_{2}^{\{1\}}\to\mathcal{K}_{2}^{\{1,2\}}$ will be added connecting $\mathcal{G}_{\{1\}}\to\mathcal{G}_{\{1,2\}}$. Similarly, edge $\mathcal{K}_{3}^{\{1\}}\to\mathcal{K}_{3}^{\{1,3\}}$ will be added connecting subgraphs $\mathcal{G}_{\{1\}}\to\mathcal{G}_{\{1,3\}}$.

\textbf{Layer $\ell.$} In general, layer $\ell$ consists of subgraphs corresponding to all subsets of $\ell$ visited tasks $B \in \mathcal{B}_\ell$, where $\mathcal{B}_\ell=\{B\subseteq\mathcal{K} \mid |B| =\ell, \mathcal{K}_1\in B\}$. Like before, a copy of $\mathcal{G}(\mathcal{V}, \mathcal{E})$ is used with a suffix applied to each task. A directed edge will be added from each parent subgraph in layer $(\ell-1)$ to the new subgraph in layer $\ell$.

\textbf{Layer $n_\mathcal{K}$.} The final layer will consist of a single subgraph, which represents all tasks visited, $\mathcal{B}_{n_\mathcal{K}}=\{\mathcal{K}\}$, with every subgraph in layer $n_{\mathcal{K}}-1$ connecting to it.

By specifying $\mathcal{S}=\mathcal{K}_{1}^{\{1\}}$ and $\mathcal{T}=\mathcal{K}_{1}^{\{1,\dots,n_\mathcal{K}\}}$ alongside the directed edges connecting the subgraphs, the AGCS has reformulated the TSP in GCS as an SPP in GCS using LTL. This now allows methods from \cite{marcucci2024shortest,marcucci2025unified,marcucci2023motion} to solve the SPP in the AGCS. A summary of the algorithm can be found in Algorithm \ref{Algo1}.

\begin{algorithm}[ht] 
    \caption{AGCS Construction}
    \begin{algorithmic}[1] \label{Algo1}
        \renewcommand{\algorithmicrequire}{\textbf{Input:}}
        \renewcommand{\algorithmicensure}{\textbf{Output:}}
        \REQUIRE GCS $\mathcal{G}(\mathcal{V}, \mathcal{E})$, source $\mathcal{S}$
        \ENSURE AGCS $\hat{\mathcal{G}}( \hat{\mathcal{V}}, \hat{\mathcal{E}})$
        \FOR {$\ell=1$ to $n_{\mathcal{K}}$}
        \STATE $\mathcal{B}_\ell = \{B \subseteq \mathcal{K} \mid |B| = \ell, \mathcal{S}\in B \} $
        \FOR {$B \in \mathcal{B}_\ell$}
        \STATE Create subgraph $\mathcal{G}_B(\mathcal{V}_B, \mathcal{E}_B)$, where $\mathcal{V}_B := \mathcal{V}$, $\mathcal{E}_B := \mathcal{E} $
        \FOR {$v \in B \setminus\{\mathcal{S}\}$}
        \STATE Add directed edge $\mathcal{E}_B \leftarrow \mathcal{E}_B \cup \{ (u, v) \} $, \\ where $u$ is the copy of $v$ in  $\mathcal{V}_{B \setminus \{ v \} }$
        \ENDFOR
        \ENDFOR
        \ENDFOR
        \RETURN $\hat{\mathcal{G}}( \hat{\mathcal{V}}, \hat{\mathcal{E}})$, where $ \hat{\mathcal{V}} = \cup_{B \in \{ \mathcal{B}_\ell \}_{\ell = 1}^{n_\mathcal{K}}} \mathcal{V}_B $, $ \hat{\mathcal{E}} = \cup_{B \in \{ \mathcal{B}_\ell \}_{\ell = 1}^{n_\mathcal{K}}} \mathcal{E}_B $ 
    \end{algorithmic}
\end{algorithm}

\begin{figure}[htbp]
  \centering
  \includegraphics[width=0.95\linewidth]{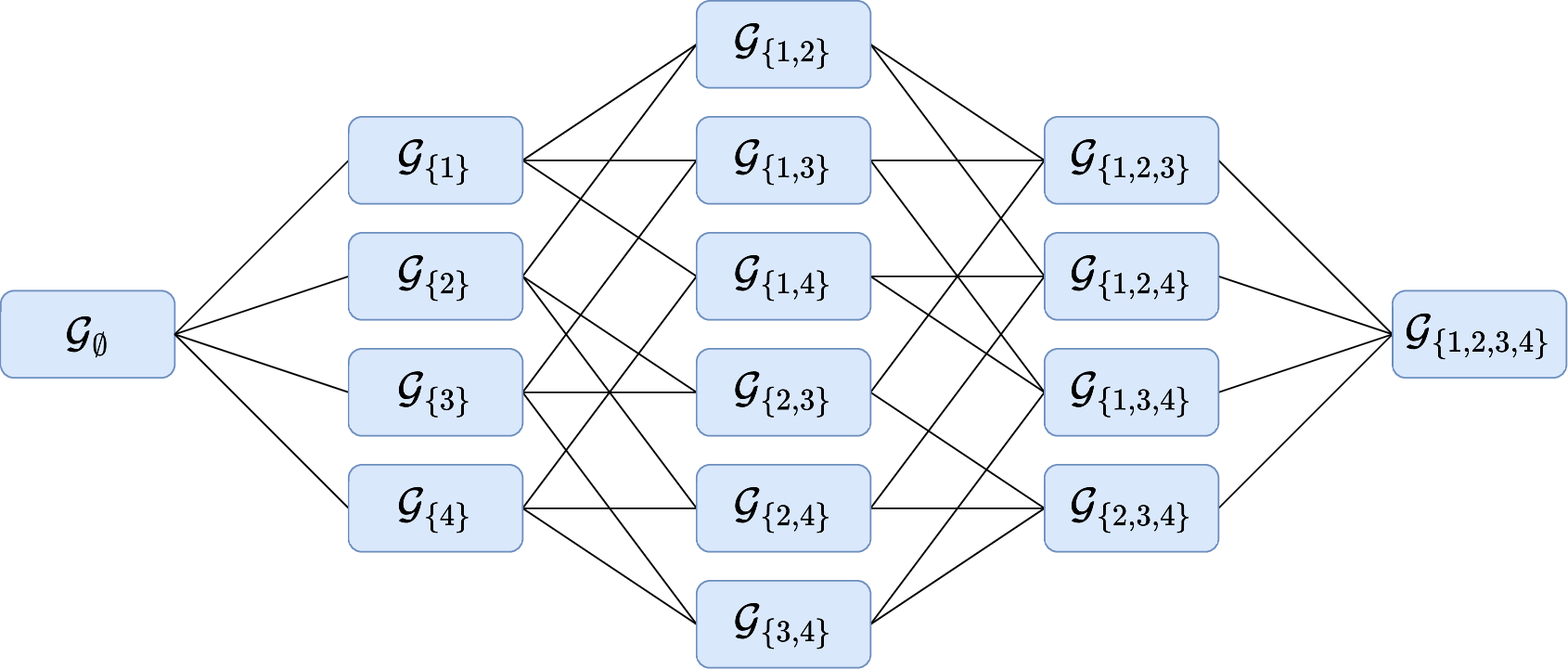}
  \caption{AGCS of a wayset-collection with four tasks. Edges between subgraphs represent the visited task. The first subgraph is $\mathcal{G}_{\emptyset}$, since the source is not a task.}
  \label{fig:agcs_4}
\end{figure}

If, instead of the TSP in GCS (Hamiltonian cycle), the application requires the wayset-collection (Hamiltonian path), such as the examples in \cite{ you2026framework}, then Algorithm \ref{Algo1} can be tweaked slightly to account for this to produce an AGCS such as the one in Figure \ref{fig:agcs_4}. Note that this produces the exact same structure as the AGCS in Figure \ref{fig:agcs_5} with the only difference being that the source and target are not considered tasks. An optimal Hamiltonian path can be found from a Hamiltonian cycle that is forced to contain the returning edge $\mathcal{T}\to \mathcal{S}$, then discarding this edge.

\subsection{Connecting AGCS and the BHK Algorithm}\label{BHK}
One core element of both the AGCS and the BHK algorithm is the use of binary subsets, $B$. These binary subsets are examined in both the AGCS and the BHK algorithm, in order of increasing size. In the AGCS, this is done using directed edges between subgraphs, which prevent backtracking in a dynamic programming style. In the BHK algorithm, this is done with a cost matrix that uses dynamic programming directly to avoid recomputing unnecessary solutions. For example, if it has been determined by the BHK algorithm that the path $1\to2\to3\to4$ is shorter than $1\to3\to2\to4$, then the path $1\to2\to3\to4\to5$ must be shorter than $1\to3\to2\to4\to5$. This behavior is explicitly encoded in the AGCS. The only difference is that this comparison is arithmetic in the BHK algorithm, but requires branch and bound in the AGCS. Both algorithms scale exponentially in the number of tasks due to the exponential number of binary subsets. This shows how the AGCS generalizes the BHK algorithm in the GCS setting.

\subsection{Properties of the AGCS}
By construction, every path in the AGCS from the source to the target satisfies the TSP in GCS or the wayset-collection. Therefore, a shortest path in the AGCS corresponds to an optimal solution of the original problem \eqref{optprob}. The mixed-integer convex reformulation of the AGCS based on \cite{marcucci2024shortest,marcucci2023motion} thus exactly solves \eqref{optprob}, up to a finite parameterization of the trajectory $q$ (e.g., using B\'ezier curves). The formulation is guaranteed to find a path if one exists.

A convex relaxation of the exact mixed-integer formulation, obtained by simply dropping the binary constraints, along with an inexpensive rounding scheme, can be used to obtain approximate solutions to \eqref{optprob}. It was observed in \cite{marcucci2023motion} that in many practical motion planning problems (without specifications), this relaxation is often tight and provides certifiably near-optimal solutions. We will study the tightness of this relaxation for problems with TSP in GCS and wayset-collection specifications in Section \ref{numeric}.

\subsection{Size of the AGCS}\label{sizeAGCS}
\begin{figure}[!t]
  \centering
  \subfloat[]{\includegraphics[width=0.45\linewidth]{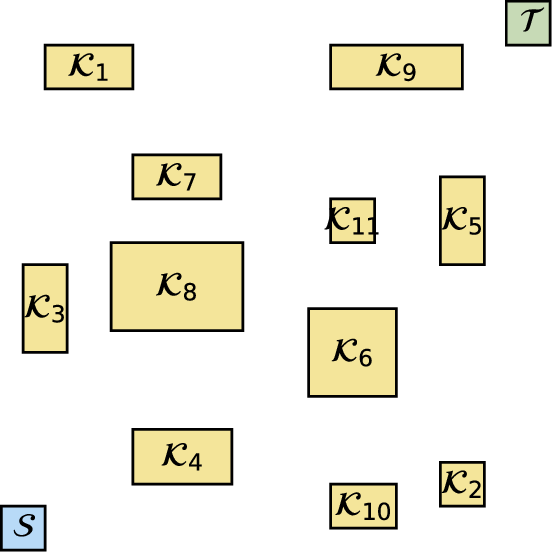}\label{fig:11_example}}
  \hfill
  \subfloat[]{\includegraphics[width=0.45\linewidth]{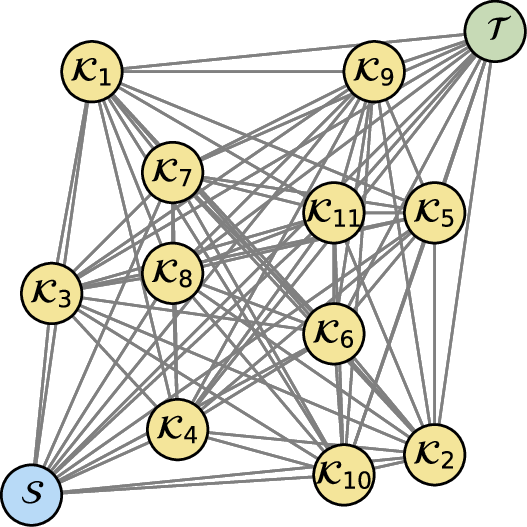}\label{fig:11_gcs}}
  \caption{(a) An 11-task example for the wayset-collection variation, blue is the source, yellow are the tasks, and green is the target. (b) GCS of the instance, a complete graph.}
  \label{fig:11_key}
\end{figure}

The number of subgraphs in the AGCS is directly correlated to the number of tasks, specifically $2^{n_{\mathcal{K}}}$. This is the same as the upper bound obtained in \cite{ you2026framework} for the AGCS with precedence specifications, where all tasks are reachable by every other task. A graph $\mathcal{G}(\mathcal{V}, \mathcal{E})$ with $|\mathcal{V}|$ vertices (tasks) will have $|\mathcal{E}| = |\mathcal{V}|\cdot(|\mathcal{V}|-1)$ edges, since each edge must be directed. In the AGCS for the TSP in GCS, $\hat{\mathcal{G}}( \hat{\mathcal{V}}, \hat{\mathcal{E}})$ will have $|\hat{\mathcal{V}}|=|\mathcal{V}|\cdot2^{|\mathcal{V}|-1}$ vertices and $|\hat{\mathcal{E}}|= (|\mathcal{V}|-1)\cdot(2|\mathcal{V}|+1)\cdot2^{|\mathcal{V}|-2}$ edges. For the wayset-collection variation with a distinct source and target, $|\hat{\mathcal{V}}|$ and $|\hat{\mathcal{E}}|$ become $|\mathcal{V}|\cdot2^{|\mathcal{V}|-2}$ and $(2|\mathcal{V}|^2-|\mathcal{V}|-2)\cdot(2^{|\mathcal{V}|-3})$, respectively. This severely limits the scale of the problems the AGCS can handle. Figure \ref{fig:11_key}, for example, will produce an AGCS with
 26\,624 vertices, 330\,752 edges, 2\,048 subgraphs, and a maximum width of 462. Such an example would be difficult to compute, motivating the use of heuristics.

\section{Pruning the AGCS}

Since the maximum width is what determines if a problem is tractable, both approaches discussed in this section aim to reduce it by either using a sparse graph or by bounding the size of the AGCS. One way to reduce the maximum width would be to remove unpromising connections in the GCS. For example, the AGCS of Figure \ref{fig:11_key} would have a connection from $\mathcal{G}_\emptyset \to \mathcal{G}_{\{9\}}$, yet this edge, $(\mathcal{S},\mathcal{K}_9)$, is highly unlikely to be part of the optimal solution. One way to remove these unpromising connections is to use a Delaunay triangulation.

\begin{remark}
Both approaches to pruning the AGCS in this section trade some optimality for tractability, so a lower bound is needed to certify how far a given solution is from optimal. We return to this in Section \ref{numeric}, where we discuss the challenges of obtaining lower bounds for the AGCS and propose one based on minimum 1-trees.
\end{remark}

\subsection{Delaunay Triangulation}\label{delaunay}
A Delaunay triangulation connects each vertex to its natural neighbors, discarding connections to distant neighbors found in a complete graph. This property has made the Delaunay triangulation a key part of many heuristics that solve the classical TSP such as in \cite{taillard2022linearithmic, tuononen2024modified}. The one challenge here is that a Delaunay triangulation requires a point, yet in general the sets will be regions, not points. This is where the Chebyshev center, the center of the largest ball that fits inside the polytope, will be used. 

\textbf{Delaunay AGCS.} Constructing the pruned AGCS that is based on a Delaunay triangulation graph requires a representative point for each set, computed as the Chebyshev center. Once these representative points are calculated, a Delaunay triangulation can be created. 

The complete AGCS is based on the $2^{n_{\mathcal{K}}}$ binary subsets that are needed to formulate this graph. Since the Delaunay triangulation will only use a portion of the binary subsets to prune the AGCS, these binary subsets need to be clearly defined. One way to obtain these binary subsets is to create an adjacency matrix of the Delaunay triangulation and, starting from a chosen source vertex, find all connected subsets. The key here is to record the parents and children of each possible subset, as this is the information needed to construct the pruned AGCS. If the source and target are distinct, then extra care is needed to ensure that the target never appears in any binary subset of size $(n_\mathcal{K}-1)$, and that every binary subset found can reach the target. With these binary subsets defined, Algorithm \ref{Algo1} can be used to construct the Delaunay AGCS.

\subsection{Window Search}

Where the Delaunay approach removes edges found in a complete graph, the window search grows the graph from an initial ordering, capping its width by construction. This approach is based on \cite{balas2000linear}, where an initial ordering is used to build permutations to solve classical TSP problems with precedence constraints. These problems are solved using dynamic programming in linear time.

\begin{figure}[htbp]
  \centering
  \includegraphics[width=0.95\linewidth]{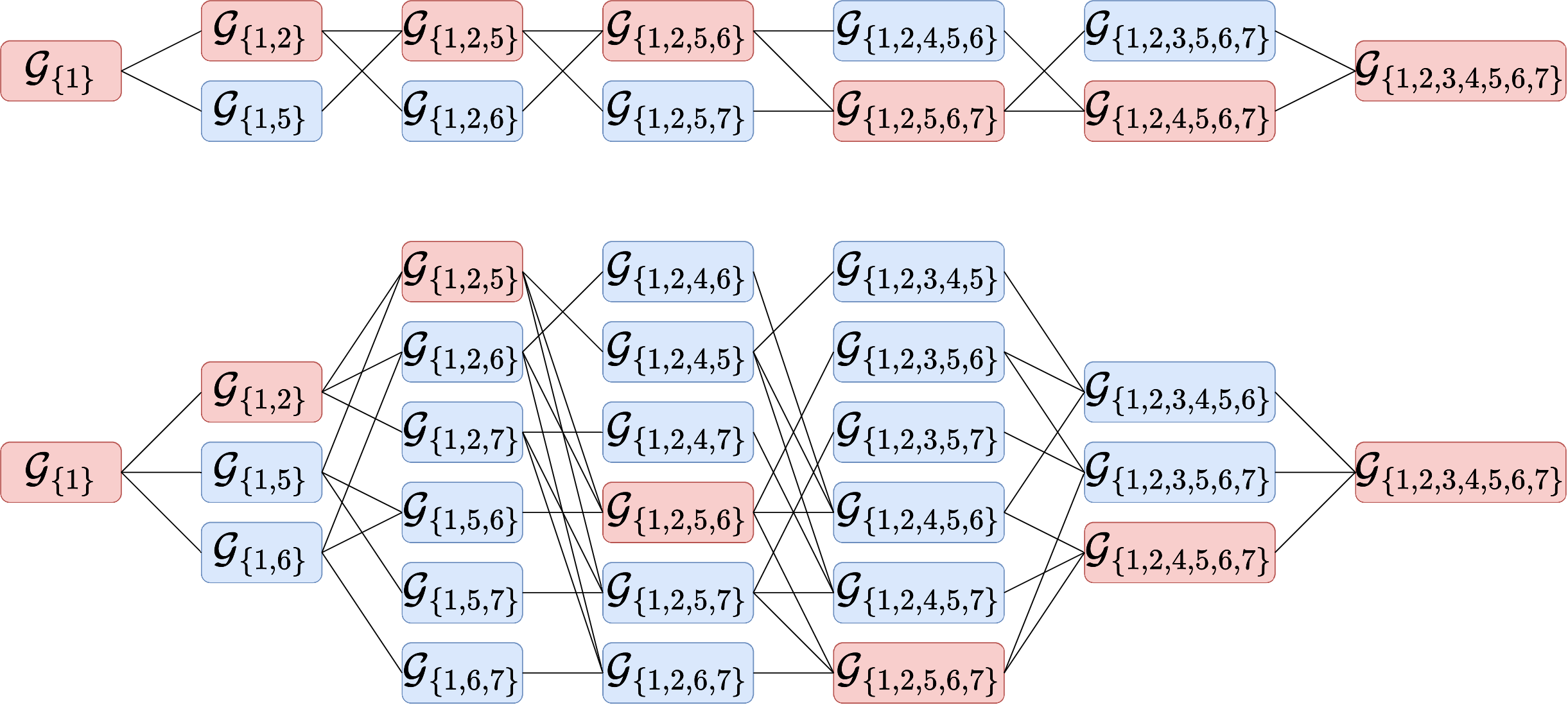}
  \caption{An initial ordering (red) being searched. (Top) AGCS with a window of one. (Bottom) AGCS with a window of two.}
  \label{fig:window}
\end{figure}

An example of a window search with windows of one and two can be seen in Figure \ref{fig:window}. Starting with the arbitrary ordering $1 \to 2 \to 5 \to 6 \to 7 \to 4 \to 3$, the window search would create the subgraphs shown in red in Figure \ref{fig:window}. On the first step, for a window search of one, 1 can search up to 5, which creates $\mathcal{G}_{\{1,5\}}$. With a window search of two, 1 can search up to 6, creating both $\mathcal{G}_{\{1,5\}}$ and $\mathcal{G}_{\{1,6\}}$. On the second step, for a window search of one, $\mathcal{G}_{\{1,2\}}$ can go to $\mathcal{G}_{\{1,2,5\}}$ or $\mathcal{G}_{\{1,2,6\}}$ since it contains 2, which was part of the original ordering. However, $\mathcal{G}_{\{1,5\}}$ can only go to $\mathcal{G}_{\{1,2,5\}}$ because 2 has not been visited. This is the key part that allows the width of the pruned AGCS to remain controlled. This same pattern continues for the rest of the steps, and is similar for a window of two.

There is a trade-off with the window size, as it increases, the search space is larger, allowing a higher chance to improve the initial ordering. However, as the window increases, so does the width, and at some point it will reach the same maximum width of the full AGCS. Therefore, the best way to take advantage of a window search is to provide a good ordering. To do so, some tools from the classical TSP will be used.

\textbf{Re-Delaunay.} As stated in Section \ref{delaunay}, the advantage of using Delaunay triangulations is that regions are connected naturally to their nearest neighbors, without doing a distance calculation. Another property of Delaunay triangulations is that they produce sparse graphs, compared to complete graphs. This reduces the number of edge cost calculations from $O(n_{\mathcal{K}}^2)$ to about $O(n_{\mathcal{K}})$. It is important to note that not all valid Delaunay triangulations are guaranteed to contain Hamiltonian cycles \cite{dillencourt1990toughness}, and finding Hamiltonian cycles in Delaunay triangulations is NP-hard \cite{dillencourt1996finding}. The Re-Delaunay heuristic \cite{tuononen2024modified} greedily selects all the shortest edges in the Delaunay graph, while maintaining a maximum degree of 2 for all vertices. Once no more valid edges can be added, the result will be disconnected paths. Another Delaunay triangulation is created from all the endpoints of all the paths, and the process continues. After a certain threshold is met, instead of calculating another triangulation, a nearest-neighbor approach connects the remaining fragments. This heuristic takes full advantage of the Delaunay graph, and overcomes its limitations by adding additional edges. The problem is that this heuristic requires an edge cost, which is not present in our problem. This is where bounded edge costs will be used.

\textbf{Bounded Edge Cost.} One approximation for edge costs is the distance between Chebyshev centers, but this can overshoot the true edge cost. Another approximate edge cost is the \emph{bounded edge cost}, which is the minimum distance between any two regions. Bounded edge costs always lower bound the true edge cost. The bounded edge cost can be calculated by solving \eqref{spp_gcs} with $\mathcal{V}_P=\{v_i,v_j\}$ and $\mathcal{E}_P=\{(v_i,v_j)\}$, with $P:=v_i\to v_j$.

\begin{figure}[htbp]
  \centering
  \includegraphics[width=1\linewidth]{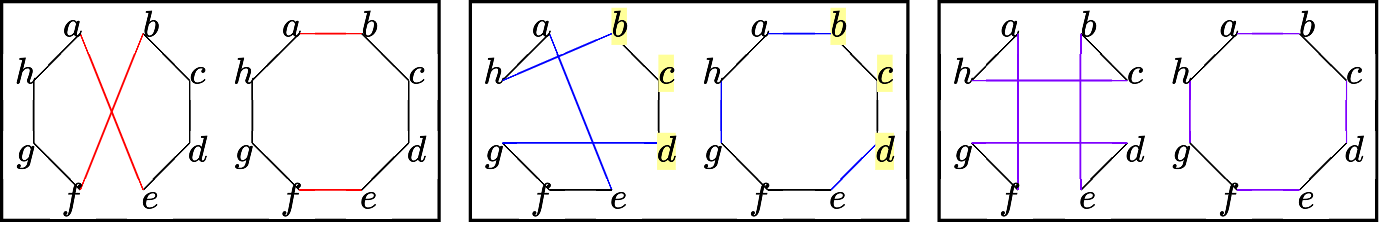}
  \caption{(Left) 2-Opt move, swapping edges $(a,e)$ and $(b,f)$ for edges $(a,b)$ and $(e,f)$. (Middle) Or-Opt move, taking the triplet $(b,c,d)$ and moving it from between $h$ and $g$ to between $a$ and $e$. (Right) Double-bridge move, swapping edges $[(a,f), (b,e),(c,h),(d,g)]$ for edges $[(a,b), (c,d), (e,f),(g,h)]$.}
  \label{fig:local}
\end{figure}

\textbf{Local Search.} One common approach to the TSP is to obtain an initial greedy tour, such as a Re-Delaunay tour, then perform local searches called $k$-Opt, which swaps $k$ edges in hopes of reducing the overall cost. These $k$-Opt heuristics improve tours by removing edges that cross over each other. In general, for $k$ edges there will be $(k-1)!\cdot2^{k-1}-1$ alternative tour combinations to check, which is why $k$ is kept small in practice. More sophisticated algorithms such as \cite{helsgaun2000effective,lin1973effective} use variable $k$-Opt, which takes candidate edge sets for each vertex and performs $k$ edge swaps, subject to these candidate edges, increasing $k$ methodically. It is important to note that these are \emph{local} searches and will often get stuck in local minima, which means only high-quality solutions are guaranteed, but optimality is not. Due to the complexity of variable $k$-Opt algorithms combined with the local minima limitation, a compromise is to use select $k$-Opt moves, specifically the 2-Opt \cite{croes1958method}, Or-Opt \cite{or1976traveling}, and the double-bridge move \cite{lin1973effective}. A 2-Opt swap is the simplest $k$-Opt move, only swapping two edges. Since it will always create exactly one alternative tour, it is also fast. An Or-Opt move is a special version of a 3-Opt move, where a select number of consecutive vertices of length one to three are placed somewhere else in the tour maintaining the same ordering. A double-bridge move is a special variant of the 4-Opt that can push other $k$-Opt moves out of local minima when performed randomly. Examples of each of these moves can be seen in Figure \ref{fig:local}. The key part that makes these local searches efficient is restricting them to a set of candidate edges rather than all possible choices. The best criterion for selecting candidate edge sets is $\alpha$-nearness using minimum 1-trees based on \cite{helsgaun2000effective}.

\textbf{Minimum 1-Trees.} Minimum (and weighted) 1-trees were first introduced in \cite{held1970traveling, held1971traveling} as a way to lower bound the classical TSP. The key idea is to select one vertex as the \emph{root}, then construct a Minimum Spanning Tree (MST) with all vertices except the root. Then the root is connected to the MST using the two shortest edges from the root. This will produce exactly one cycle in the tree. Removing one vertex from a tour will produce a tree, and adding back this vertex will produce a cycle. This is why a minimum 1-tree lower bounds the TSP. If edge costs are modified such that $$\bar{c}_{ij}=c_{ij}+\pi_i+\pi_j$$ where $\bar{c}_{ij}$ is the modified cost, $c_{ij}$ is the original cost, and $\pi_i$ is the penalty associated with vertex $i$, then this will produce a \emph{weighted} 1-tree. These penalties $\pi$ can be calculated based on the degree of vertex $i$: if it is less than two, the penalty decreases the associated edge costs, and if it is more than two then the penalty increases them. Using an ascent method, this will incentivize the weighted 1-tree to take the form of a cycle, providing a better lower bound for the classical TSP. The true lower bound cost can be obtained by subtracting $2\sum_{i=1}^{n}\pi_i$ from the weighted 1-tree cost \cite{held1970traveling, held1971traveling}.

\textbf{Candidate Edge Set.} Most of the edges in the weighted 1-tree will most likely also be part of the solution to the TSP. This is what makes them excellent edges to fill the candidate edge set for each vertex. In \cite{helsgaun2000effective}, the metric $\alpha$-nearness is defined as the additional cost added to the weighted 1-tree if edge $(i,j)$ were to be forced into the weighted 1-tree. This is proven to be a better metric for creating a candidate edge set compared to ranking edges purely on a nearest-neighbor basis. Candidate sets also pair naturally with sparse cost matrices. A Delaunay graph already excludes unpromising edges, and ranking them by $\alpha$-nearness further improves efficiency.

\textbf{Initial Ordering.} With all the different parts laid out, an ordering can now be created for the windowed search. First, a Delaunay triangulation is calculated based on the Chebyshev centers of each region. Using the edges from the Delaunay graph $\mathcal{D}(\mathcal{{V}}, \mathcal{\bar{E}})$, a sparse cost matrix, $c_{ij}$, is created using bounded edge costs. This sparse cost matrix is then used for the Re-Delaunay heuristic to obtain an initial ordering, $\sigma_0$. After the initial ordering is set, a candidate edge set, $\mathcal{C}_i$, is produced based on $\alpha$-nearness for every vertex. This is then used for an iterative local search composed of a double-bridge move followed by 2-Opt and Or-Opt re-optimizations, to improve the initial ordering. The final step is to perform a windowed search on this improved ordering, $\sigma$, to build a pruned AGCS, $\check{\mathcal{G}}( \check{\mathcal{V}}, \check{\mathcal{E}})$, to be solved. Algorithm \ref{Algo2} summarizes the heuristic.

\begin{algorithm}[ht] 
    \caption{Window Search Heuristic}
    \begin{algorithmic}[1] \label{Algo2}
        \renewcommand{\algorithmicrequire}{\textbf{Input:}}
        \renewcommand{\algorithmicensure}{\textbf{Output:}}
        \REQUIRE GCS $\mathcal{G}(\mathcal{V}, \mathcal{E})$, window size: $W$
        \ENSURE Pruned AGCS $\check{\mathcal{G}}( \check{\mathcal{V}}, \check{\mathcal{E}})$
        \STATE Create Delaunay Graph, $\mathcal{D}(\mathcal{{V}}, \mathcal{\bar{E}})$
        \STATE Get cost matrix, $c_{ij}$ $\leftarrow$ \texttt{cost$\left(\mathcal{D}(\mathcal{{V}}, \mathcal{\bar{E}})\right)$}
        \STATE Initial ordering, $\sigma_0 \leftarrow$ \texttt{re\_delaunay$\left(c_{ij}\right)$}
        \STATE Get candidate edge set, $\mathcal{C}_i \leftarrow$ \texttt{candidates$\left(c_{ij}\right)$}
        \STATE Perform local search, $\sigma \leftarrow$ \texttt{improve$\left(\sigma_0,c_{ij},\mathcal{C}_i\right)$}
        \STATE Obtain pruned AGCS, $\check{\mathcal{G}}( \check{\mathcal{V}}, \check{\mathcal{E}}) \leftarrow$ \texttt{window$\left(\sigma, W\right)$}
        \RETURN $\check{\mathcal{G}}( \check{\mathcal{V}}, \check{\mathcal{E}})$
    \end{algorithmic}
\end{algorithm}

\section{Numerical Experiments}\label{numeric}
We compare both methods against the full AGCS on a small instance where the optimum is known, then push the window search to instances far beyond the reach of the Delaunay and full AGCS. For all examples, the Euclidean distance was used as the edge cost function, a simplification of the general convex costs in \eqref{optprob}. All experiments were performed on a laptop with an Intel Core Ultra 7 258V and 32 GB of RAM. Experiments were performed using Drake \cite{drake} and MOSEK \cite{mosek}.

\begin{figure}[htbp]
  \centering
  \includegraphics[width=0.75\linewidth]{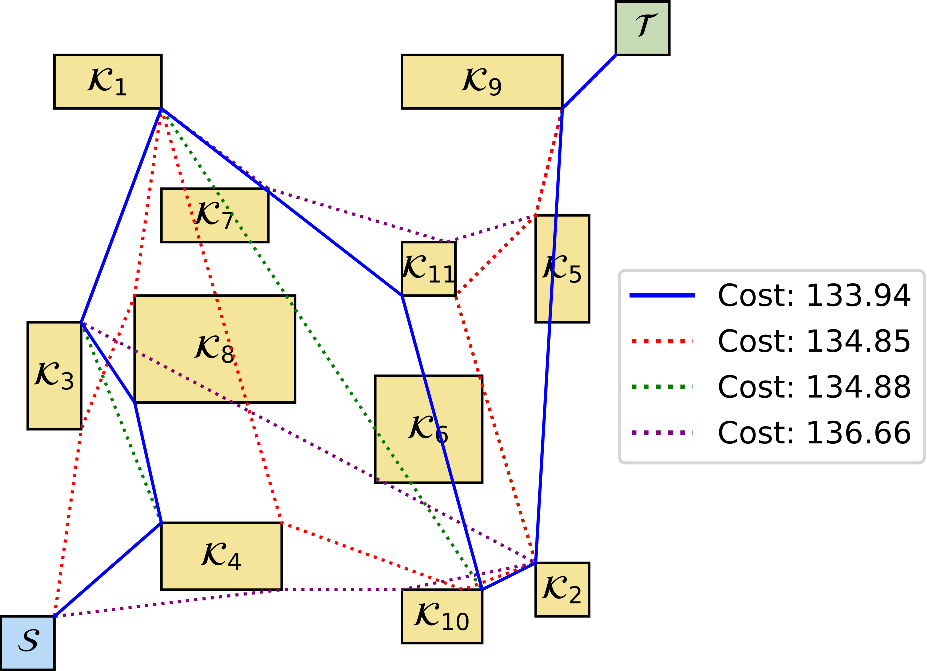}
  \caption{All distinct solutions for example in Figure \ref{fig:11_key}. Solid blue line is the certified optimal solution.}
  \label{fig:all_sol}
\end{figure}

The first experiment will be a comparison of the different algorithms on the instance in Figure \ref{fig:11_key}. The results can be seen in Figure \ref{fig:all_sol}, which compares the cost of the complete AGCS, the Delaunay AGCS, and a windowed heuristic with $W=\{0,\dots,6\}$. Table \ref{tab:agcs_comp} shows further details on the number of subgraphs $|\hat{\mathcal{G}}|$, maximum width $\bar{w}$, solution cost $C$, optimality gap $\delta$, and solve time $\tau$.

\begin{table}[htbp]
\centering
\caption{Comparison of AGCS, Delaunay AGCS, and various window searches for the example in Figure \ref{fig:11_key}}
\label{tab:agcs_comp}
\begin{tabular}{l c c c r r}
\toprule
Type & $|\hat{\mathcal{G}}|$ & $\bar{w}$ & $C$ & $\delta$ (\%) & $\tau$ (s) \\
\midrule
AGCS & 2048 & 462 & 133.94 & 0.00 & 801.52 \\
Delaunay & 1193 & 321 & 133.94 & 0.00 & 98.10 \\
$W=6$ & 1585 & 462 & 133.94 & 0.00 & 232.34 \\
$W=5$ & 925 & 252 & 134.85 & 0.68 & 74.24 \\
$W=4$ & 393 & 70 & 134.88 & 0.70 & 18.69 \\
$W=3$ & 151 & 20 & 136.66 & 2.03 & 3.73 \\
$W=2$ & 57 & 6 & 136.66 & 2.03 & 0.67 \\
$W=1$ & 23 & 2 & 136.66 & 2.03 & 0.19 \\
$W=0$ & 13 & 1 & 136.66 & 2.03 & 0.11 \\
\bottomrule
\end{tabular}
\end{table}

We compared the effects of changing the window, $W$, for the window search heuristic on TSP in GCS instances of varying sizes. These randomly generated instances consist of $n_{\mathcal{K}}$ task regions placed uniformly on an $n_{\mathcal{K}}\times n_{\mathcal{K}}$ grid in $[0,1]^2$. For all instances, $W=0$ is solved first, then the largest $W$ that can be calculated in one minute is reported. The results can be seen in Table \ref{tab:tsp_window}, where $C_0$ and $\tau_0$ are the cost and solve time for $W=0$, and $C$ and $\tau$ are the cost and solve time for the specified $W$. The number of subgraphs $|\hat{\mathcal{G}}|$ and the maximum width $\bar{w}$ are also stated. Note the $n_\mathcal{K}=1000$ in Figure \ref{fig:large_tsp} and the one in Table \ref{tab:tsp_window} are different.

\begin{table}[htbp]
\centering
\caption{Comparison of $W$ with various TSP in GCS instances.}
\label{tab:tsp_window}
\begin{tabular}{l r r c r r c c}
\toprule
$n_{\mathcal{K}}$ & $C_0$& $\tau_0$ (s) & $W$ & $C$ & $\tau$ (s) & $|\hat{\mathcal{G}}|$ & $\bar{w}$\\
\midrule
20 & 2.73 & 0.10 & 4 & 2.73 & 59.68 & 952 & 70\\
60 & 4.72 & 0.29 & 3 & 4.72 & 39.51 & 1110 & 20\\
100 & 5.51 & 0.45 & 2 & 5.51 & 11.00 & 584 & 6\\
200 & 7.40 & 0.91 & 2 & 7.40 & 27.36 & 1184 & 6\\
600 & 12.85 & 2.64 & 1 & 12.82 & 14.36 & 1198 & 2\\
1000 & 16.81 & 4.09 & 1 & 16.80 & 21.71 & 1998 & 2\\
\bottomrule
\end{tabular}
\end{table}

\textbf{Discussion.} From Figure \ref{fig:all_sol} and Table \ref{tab:agcs_comp}, both the Delaunay AGCS and the pruned AGCS with $W=6$ reach the optimum faster than the full AGCS. Also, smaller windows trade a 0.68--2.03\% gap for one to four orders of magnitude in solve time. From Table \ref{tab:tsp_window}, at larger sizes even $W=1$ improves the tour cost. The unchanged costs at smaller sizes do not indicate optimality, rather the window was too narrow to find any improved reordering. This can be seen in Table \ref{tab:agcs_comp} where the cost remains at 136.66 until $W=4$. 

In conclusion, the Delaunay AGCS can replace the full AGCS on small instances. The window search extends to instances beyond the reach of either. Note from Table \ref{tab:tsp_window} that the maximum width stays bounded even as $n_{\mathcal{K}}$ increases. The methods presented here create a good foundation for pruning the AGCS. Improvements in the initial ordering could provide better and faster solutions.

\begin{figure}[!ht]
  \centering
  \subfloat[]{\includegraphics[width=0.3\linewidth]{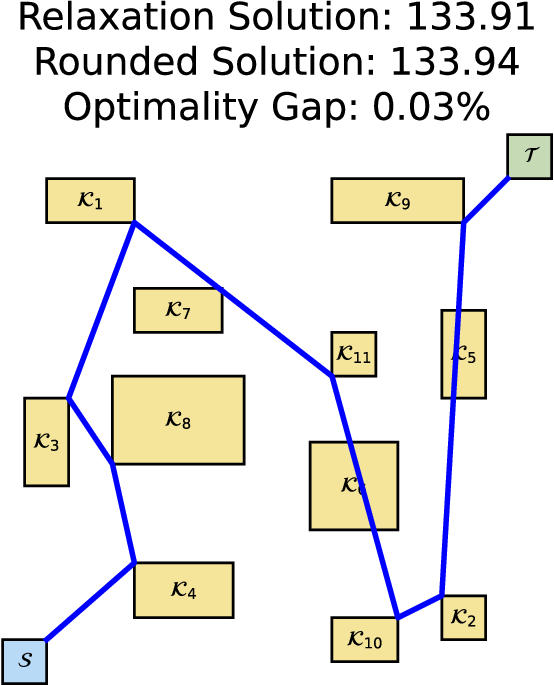}\label{fig:unbounded}}
  \hfill
  \subfloat[]{\includegraphics[width=0.3\linewidth]{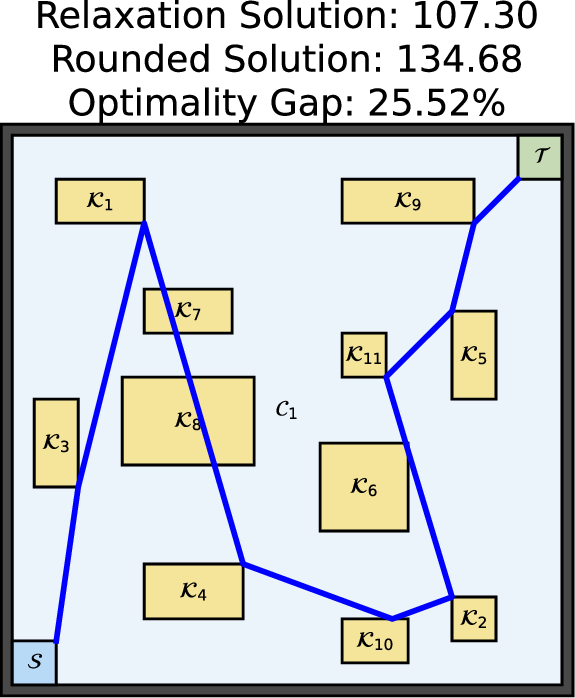}\label{fig:bounded}}
  \hfill
  \subfloat[]{\includegraphics[width=0.3\linewidth]{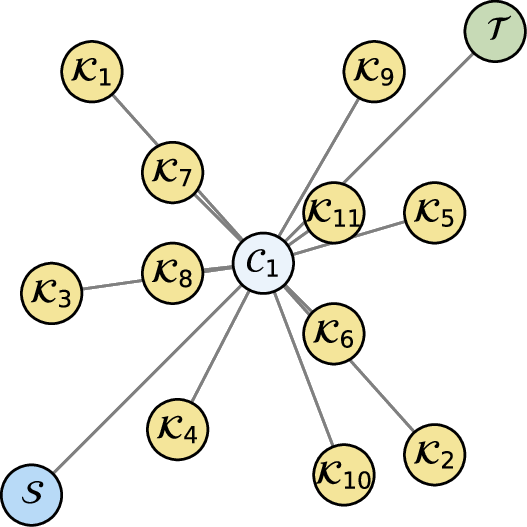}\label{fig:bounded_gcs}}
  \caption{(a) Relaxation and best rounded solution for the instance in Figure \ref{fig:11_key}, showing a certified optimal solution. (b) Relaxation and best rounded solution for the same instance, placed in a bounded environment. (c) GCS of the bounded instance.}
  \label{fig:relaxation}
\end{figure}

\textbf{Lower Bounds.} For the AGCS, certifying solution quality means comparing against the convex relaxation, which can itself be loose. In Figure \ref{fig:relaxation}, the instance from Figure \ref{fig:11_key} in an unbounded environment with a complete GCS yields a certified optimal solution. The same instance in a bounded environment with a sparse GCS produces both a worse relaxation and a sub-optimal rounding, reporting a 25.52\% gap when the true gap is 0.55\%.

One takeaway from this is that lower bounds need the connectivity between tasks such as the one found in Figure \ref{fig:11_gcs}, rather than the connectivity from Figure \ref{fig:bounded_gcs}. Another important note regarding unbounded and bounded environments is that bounded environments, with more obstacles, will also have more trivial solutions. This is because obstacles will naturally force certain orderings, creating subproblems that can be solved separately, such as the one in Figure \ref{fig:traj}. 

Another challenge is that the convex relaxation of the AGCS is solved using branch and bound, meaning the relaxation itself will not scale well. This means formulations for the MST and TSP in GCS, such as the mixed-integer program ones found in \cite{marcucci2025unified}, would suffer the same limitations, since they too require branch and bound.

The fastest and most scalable way to obtain a lower bound would be to use the bounded edge costs and solve a weighted 1-tree. Since the bounded edge costs themselves lower bound true edge costs and the weighted 1-tree lower bounds the TSP, this is currently the best way to guarantee a lower bound on any TSP in GCS or wayset-collection instance. However, the lower bound can remain loose even if the achieved solution is already optimal. The gap arises because bounded edge costs allow each region to be represented by a different point on each incident edge, whereas a feasible trajectory must choose a single point per region. This gap is a property of geometry, not of the search. Using this conservative lower bound, the example in Figure \ref{fig:large_tsp} has a gap of 79.71\%. Table \ref{tab:lower_bound} shows a few examples of the optimality gap of certified optimal solutions and the proposed lower bound. Note that computing lower bounds requires a complete cost matrix. Although these gaps are loose, no other tractable method currently produces a guaranteed lower bound for large-scale instances of the TSP in GCS. Future work will explore ways to obtain tighter lower bounds for the TSP in GCS and wayset-collection.

\begin{table}[htbp]
\centering
\caption{Gap comparison between certified optimal solution and the proposed lower bound, 100 trials for each.}
\label{tab:lower_bound}
\begin{tabular}{l r r r r}
\toprule
$n_{\mathcal{K}}$ & mean (\%) & median (\%) & min (\%) & max (\%)\\
\midrule
5 & 10.54 & 9.96 & 2.39 & 29.14 \\
7 & 15.75 & 15.51 & 7.10 & 34.30 \\
9 & 17.11 & 16.77 & 7.81 & 33.27\\
11 & 17.81 & 16.60 & 10.41 & 33.86\\
\bottomrule
\end{tabular}
\end{table}

\bibliographystyle{IEEEtran}
\bibliography{bibliography.bib}

\end{document}